# Diversity in News Recommendation

Abraham Bernstein[*,1], Claes de Vreese[*,2], Natali Helberger[*,3], Wolfgang Schulz[*,4], Katharina Zweig[*,5], Christian Baden[*,6], Michael A. Beam[7], Marc P. Hauer[8], Lucien Heitz[†,9], Pascal Jürgens[10], Christian Katzenbach[11], Benjamin Kille[12], Beate Klimkiewicz[13], Wiebke Loosen[14], Judith Moeller[15], Goran Radanovic[16], Guy Shani[17], Nava Tintarev[18], Suzanne Tolmeijer[†,19], Wouter van Atteveldt[20], Sanne Vrijenhoek[21], and Theresa Zueger[22]

1   Universität Zürich, CH. bernstein@ifi.uzh.ch
2   University of Amsterdam, NL. c.h.devreese@uva.nl
3   University of Amsterdam, NL. n.helberger@uva.nl
4   Universität Hamburg, DE. w.schulz@hans-bredow-institut.de
5   Technische Universität Kaiserslautern, DE. zweig@cs.uni-kl.de
6   The Hebrew University of Jerusalem, IL. c.baden@mail.huji.ac.il
7   Kent State University, US. mbeam6@kent.edu
8   Technische Universität Kaiserslautern, DE. hauer@cs.uni-kl.de
9   Universität Zürich, CH. heitz@ifi.uzh.ch
10  Johannes Gutenberg-Universität Mainz, DE. pascal.juergens@uni-mainz.de
11  Alexander von Humboldt Institute for Internet and Society, DE. katzenbach@hiig.de
12  Technische Universität Berlin, DE. benjamin.kille@tu-berlin.de
13  University Jagiellonski, PL. beatakl@hotmail.com
14  Leibniz Institute for Media Research, Hans-Bredow-Institut, DE. w.loosen@leibniz-hbi.de
15  University of Amsterdam, NL. j.moller@uva.nl
16  Max Planck Institute for Software Systems, DE. gradanovic@mpi-sws.org
17  Ben Gurion University, IL. shanigu@bgu.ac.il
18  Maastricht University, NL. n.tintarev@maastrichtuniversity.nl
19  Universität Zürich, CH. tolmeijer@ifi.uzh.ch
20  Vrije Universiteit Amsterdam, NL. w.h.van.atteveldt@vu.nl
21  University of Amsterdam, NL. s.vrijenhoek@uva.nl
22  Alexander von Humboldt Institute for Internet and Society, DE. zueger@hiig.de

## Abstract

News diversity in the media has for a long time been a foundational and uncontested basis for ensuring that the communicative needs of individuals and society at large are met. Today, people increasingly rely on online content and recommender systems to consume information challenging the traditional concept of news diversity. In addition, the very concept of diversity, which differs between disciplines, will need to be re-evaluated requiring an interdisciplinary investigation, which requires a new level of mutual cooperation between computer scientists, social scientists, and legal scholars. Based on the outcome of a interdisciplinary workshop, we have the following recommendations, directed at researchers, funders, legislators, regulators, and the media industry:
- Conduct interdisciplinary research on news recommenders and diversity.
- Create a safe harbor for academic research with industry data.
- Strengthen the role of public values in news recommenders.
- Create a meaningful governance framework for news recommenders.
- Fund a joint lab to spearhead the needed interdisciplinary research, boost practical innovation, develop reference solutions, and transfer insights into practice.

---

\* Editor / Organizer
† Editorial Assistant / Collector

**Perspectives Workshop** November 24–29, 2019 – www.dagstuhl.de/19482
**2012 ACM Subject Classification** Information systems → Web services; Information systems → Information retrieval diversity; Applied computing → Psychology; Human-centered computing → Empirical studies in HCI; Applied computing → Sociology; Information systems → Digital libraries and archives; Human-centered computing → HCI theory, concepts and models; Applied computing → Economics
**Digital Object Identifier** 10.4230/DagMan.9.1.43
**Funding** Abraham Bernstein, Lucien, Heitz, and Suzanne Tolmeijer would like to thank the Haslerstiftung for their generous support of their their reserach on news recommender systems.
**Acknowledgements** The participants wish to express their gratitude to Schloss Dagstuhl for their strong support and hosting of this workshop. The editors want to thank all participants for their contributions to the Dagstuhl Perspectives Workshop Manifesto.

## Executive summary

News diversity in the media has for a long time been a foundational and uncontested basis for ensuring that the communicative needs of individuals and society at large are met. Today, people increasingly rely on online content and recommender systems to consume information, engage in debates, and form their political opinions, as well as find relevant information in the ever-expanding information sphere. These fundamental changes in the use of content and the role of recommender systems to curate access to news challenge the realization of the traditional concept of news diversity. News diversity also becomes a matter of diversity in recommendations.

In the social sciences, and in media law and policy, the concept of diversity has always been dependent on the context of its use (e.g., dependent on which theory of democracy is followed). To develop an updated notion of diversity that takes into account the technological changes, we need to go back to the functions (or contexts) that news diversity fulfills for society. Such a functional notion of diversity will require reevaluation of core societal values, and require interdisciplinary[1] research. This interdisciplinarity will then require a new level of mutual cooperation between computer scientists, social scientists – in particular from communication science, but also from psychology, sociology, political philosophy, and economics – and legal scholars. The traditional approaches to cooperation, which rely on dividing the problem into its disciplinary parts or opportunistically regard the other disciplinary as a service provider, will not suffice anymore.

**Recommendations.** Based on the outcome of an interdisciplinary workshop, we have the following recommendations, directed at researchers, funders, legislators, regulators, and the media industry:
1. *Conduct interdisciplinary research on news recommenders and diversity*: As most pressing societal and scholarly questions about news recommender systems and diversity cannot be answered meaningfully from a mono-disciplinary perspective, we call upon the (inter)national research community to organize and engage in truly interdisciplinary, continuously cooperating communities across computer, social, and legal sciences.

---

[1] We define interdisciplinary research as "any study or group of studies undertaken by scholars from two or more distinct scientific disciplines. The research is based upon a conceptual model that links or integrates theoretical frameworks from those disciplines, uses study design and methodology that is not limited to any one field, and requires the use of perspectives and skills of the involved disciplines throughout multiple phases of the research process." [1, p. 341].

2. *Create a safe harbor for academic research with industry data*: Much research on public communication and recommender systems requires access to industry data to produce results that are meaningful for society. To enable this, data protection issues must be resolved. We recommend creating a code of conduct under Article 40 of the General Data Protection Regulation (GDPR) to give this kind of data sharing a solid legal basis.
3. *Strengthen the role of public values in news recommenders*: News recommenders can be powerful tools to help users find their way in the plethora of available news, shape public opinion, and serve as a foundation for public cohesion. They are extensions of the traditional editorial task. Hence, we recommend that they should not just maximize for clicks and short-term revenue, but, mindful of the democratic function of the media, also strengthen public values that align with the overall mission of a news outlet.
4. *Create a meaningful governance framework for news recommenders*: While we see no fruitful way of transferring existing regulations from broadcasting to news recommenders, we recommend that regulators and legislators support the research required to build diversity-aware recommender systems and actively foster an environment that allows for the co-existence of multiple recommender systems and their preconditions. Such initiatives should be evidence-based.
5. *We recommend founding a joint labs to spearhead the needed interdisciplinary research, boost practical innovation, develop reference solutions, and transfer insights into practice.* This initiative and its lab must combine the best (inter)national expertise from fields such as computer science, social and behavioral sciences, political philosophy, and law, as well as industry and regulators to ensure diverse, transparent, explainable, and fair news recommendations.

## Table of Contents

## 1   Introduction

News is a specific kind of information with respect to its unique function in society. However, what different actors consider to be news depends on a variety of patterns of interpretation. The social and societal function of "news" is that it should enable citizens to know what is regarded as important, contested, or an issue of public interest that should be deliberated upon. News shapes how people relate to their communities and society, and how they form their opinions about public affairs. News fundamentally contributes to the construction of public spheres, which need to be open to all topics, events, and opinions that a society needs for its self-observation. This makes news a crucial part of various social communication processes.

The *diversity* of news fulfills an important function for this process, displaying the variety of opinions in society to enable the emergence of shared knowledge about current affairs. This facilitates public deliberation and the creation of social cohesion. Hence, diversity is crucial for free and open public and individual opinion formation in democratic societies.

Given the amount of information, meaningful selection has become more crucial than ever, both for those who produce news and for those who use it. Selection decisions were traditionally made by human editors, but there is an emerging trend of automated news recommenders that serve similar goals. These technical systems aggregate, filter, select, and prioritize information. In doing so, they take on a powerful gatekeeping function in the information ecosystem.

A news recommender system makes automated decisions as to which news content will be presented to users. The value of these systems lies in their ability to adjust decisions to the different information needs and preferences of users (*personalization*). Such systems may make editorial decisions for the front page of an online newspaper, decide which additional stories to suggest to a reader of a story, decide on the selection and order of stories in the consumption flow of the user, unlock the long tail, and increase coverage. They are also routinely used by human editors as decision support systems. As news recommenders now influence the way in which many people are both exposed to and consume news, it is critical to understand and take into account how such systems affect the diversity of news consumption and exposure. To uncover whether diversity is, or can be, formalized sufficiently to address its societal function, we need to rethink diversity within the role of the media in a democratic society. This includes reflecting on what we 'optimize' with respect to diversity within recommender systems.

These questions are all the more urgent as different organizations and/or stakeholders have emerged that contribute to the flow of news and the use of recommender systems for part of the news selection and distribution process, including:

1. **Media organizations**, such as online newspapers, often make use of features of recommender systems, such as a list of additional stories suggested within a given story or a personalized front page.
2. **News aggregators**, such as Google News or Apple News, do not generate their own news content, but gather content from a variety of different sources, mostly journalistic, and select which stories, and in what order, to present to a given user.
3. **Social media platforms**, such as YouTube or Facebook, typically present personalized streams of content to users. These streams often include news but also a variety of different types of content.

Realizing diversity in news recommendations takes place in the dynamic interplay between these stakeholders, and the growing influence of news recommenders and the automation of editorial decisions. This work requires an interdisciplinary effort and discussion between social scientists, legal scientists, and computer scientists, as well as industry and regulators. Only together can we identify goals and solutions that are both important and realizable, as well as economically viable. The shared goal must be to ground diversity in various ways: socially and computationally. To best address diversity in news recommenders, we need to bring together researchers from different disciplines and engage them in meaningful collaboration.

The importance of a collaboration between the different fields becomes especially evident when looking at the current problems posed by diversity in news recommendations. There is currently no systematic match of research. Instead, we see the involved disciplines tackling the problem of diversity in isolation. Hence, solutions to the challenges posed by diversity in news recommenders are vague and dispersed throughout the literature.

In computer science, diversity-optimization is a research area that has a long history [2] – accordingly, its definitions are manifold [30]. In this context, the limiting factor is that diversity in computer science is mainly operationalized in the context of commercial systems [44, 28], ignoring the social element that diversity entails [21]. The discussion of what an appropriate definition of diversity is and how to measure is rooted in a normative definition. It is a non-technical question that calls for a cooperation of computer science and social science. In the interplay of disciplines, the role of computer science is to explain and inform other disciplines about technical possibilities and how to best set up and design the next generation of recommender system.

While the field of computer science has knowledge about how recommender systems operate, the social sciences and communication sciences have longstanding experience with research explaining why diversity is important, and what the effect of exposure to diverse content is on users and society. Different strands of research can be discerned: theoretical research in e.g., political philosophy, political sciences, or media studies uncovers the political underpinnings of diversity as a normative concept [4, 38, 22], respectively ways of conceptualising and monitoring diversity [27, 41]. Research in media law is studying the optimal conditions for realizing media diversity and protecting diversity as a public value as well as element of freedom of expression [6, 39, 35], also in the light of the growing proliferation of digital technology and digital platforms. Empirical research, e.g., in the communication sciences and journalism studies add to our knowledge of the effects of exposing users to diverse contents, respectively how the use of digital recommenders can affect the realisation of media diversity and the role of the media in a democratic society [19, 43, 11, 8, 18, 15, 10]. Research in the area of psychology also acknowledges the importance of diversity [17] and found that, for example, exposure to both viewpoint diversity and background diversity have a positive influence on critical thinking and problem solving [16, 12, 13].

Despite the seemingly obvious advantages of cooperation between the different strands of research there so far has been little collaboration on diversity in news recommendations [22, 23, 31]. Clearly, there is much more to be gained by supporting and encouraging the collaboration between different research fields, regulators, system designers, as well as other parties, such as media practitioners and digital platforms.

The remainder of this manifesto develops a set of recommendations for researchers in academia and industry, regulators, funders, and media organizations. It roots these recommendations in the importance of diversity as a societal concept and a means to the realization of democratic values. The manifesto considers the social and user perspective on diversity, complexities that arise from data access, design, and measurement, as well as challenges that arise in the governance of diversity in media. It closes by distilling the key issues into five core recommendations, hereby calling for a new initiative on research on news recommendations.

> **The use of recommender systems for news vs. other digital media**
>
> Today, recommender systems area part of many applications, such as movie streaming services (e.g., Netflix), e-commerce websites (e.g., Amazon), and social networks (e.g., Facebook). News recommender systems are substantially different from recommender systems in other domains. For example, the content flow in news is substantially faster than, for example, in the movie domain, while the metadata associated with news is often less detailed than, for example, the technical specification of an electronic gadget. In addition, the role of diversity in news is very different to the role of diversity elsewhere.
>
> In entertainment recommendations, the goal of the system is to help the user to rapidly find a relevant item to consume. The system is typically indifferent as it optimizes the goal of a company, which might entail an aspect concerning the utility of the platform to its users, where neither needs to be aligned to societal goals. For example, in the movie domain, YouTube's goal was to maximize the number of hours watched per day by making "relevant" movie recommendations to their users. By setting the focus on relevance, diversity is mainly considered in terms of topic diversity and/or to compensate for the inaccuracy in relevance estimation.
>
> Another common domain for recommender systems is in e-commerce applications, such as in electronic gadgets or an online supermarket. In these domains, diversity is arguably even less important. On the other hand, there may be other domains, such as court decisions recommendations, where diversity can be extremely important, but can take on a different meaning than for the news sector. These examples illustrate that diversity should always be evaluated with regards to the specific domain in question.

## 2   Why is diversity important?

In a democracy, an informed citizenship is a critical precondition, as is the availability of a public forum where the different ideas and opinions in a democratic society can be articulated, encountered, debated, and weighed [6, 4, 40]. This role (or function) has not changed and it becomes even more important in the digital realm (see Section 1). The importance of diversity extends to the way media engage with digital technologies, including the increasing automation of communicative processes [7]. In light of this role, diversity is a means to an end: to further the goals and values diverse news content helps to promote. Exposure to diverse news is a precondition for social cohesion, tolerance, and peaceful coexistence of different cultures, ideologies, and viewpoints [14, 24, 29].

Diversity of news is deeply ingrained in our understanding of what it means to live in a democratic society – a society that embraces the idea that each member of a democracy is entitled to a set of fundamental rights, including political rights, and is able to participate and have a voice. In a diverse media landscape, one actor should not be able to dominate public discourse [33]. Diversity is an inherently normative concept [38], but at a most basic level, one could define diversity as the "heterogeneity of media content in terms of one or more specified characteristics" [42]. The concrete goals and values that a diverse news recommender should optimize for, hence, depend on the understanding of democracy and the role of journalism [22, 38]. The way media engage with algorithmic recommendations must be driven by the sense of the responsibility that comes with the power of recommenders to steer reading behavior, combined with an understanding of how different measures of diversity are conducive to different functions and democratic values.

In designing both new recommender systems as well as the means to evaluate and regulate such systems, it is important to accept that diversity and democratic values are complex, diffuse, and sometimes not directly measurable or quantifiable. Despite – or maybe exactly because of – its pivotal role in society, there are different aspects of diversity that need to be taken into account, like diversity of topics, perspectives, and actors, but also language, style of news, and audiences to reach [5]. The relative importance of each dimension can be derived from the desired societal values that drive diversity. Not all of these dimensions of diversity are readily operationalizable [2] and measurable. Neither is there a single gold standard measure of diversity. Instead, the type of diversity that a media organization adheres to depends on its editorial mission, its business model, and the role of recommendations in this context [22]. Concretely, this may lead to different forms of diversity in recommendations for social media platforms and legacy media.

Furthermore, even if a way of measuring diversity is agreed upon, uncritically maximizing an arbitrary measure of diversity can result in a dysfunctional fragmentation and polarization of the news supply: the actual effect of diversification is contingent on the content and the user as well as on the desired functional values. Uncritically maximizing diversity also ignores the fact that diversity needs to be balanced and aligned with the economic interests of publishers, and user needs, including privacy, fairness, and inclusiveness [32]. The desired degree of diversity will differ depending on the relative importance of, for example, informing, fostering discussion, and representing minority voices.

Finally, diversity is not only about including different ideas and perspectives, etc. but also about making purposeful choices. The role of a recommender should be to guide news consumption in an increasingly fragmented digital environment. This is why it is not sufficient to formalize diversity in recommendations. We also need to be mindful of the way people interact with recommendations and the degree of agency publishers and designers of recommender systems share with users (see Section 3). Diverse recommendations are furthermore modulated by their integration into the broader organizational context of a newsroom, and the division of work and responsibilities between journalists, editors, and computer scientists or software developers. This is why, ultimately, diversity in recommendations is not only a matter of the design of the recommendation algorithms and the values it is being optimized for, but also of the way newsrooms embrace and understand recommendations as an extension of their editorial task.

When it comes to analyzing the societal effects of diversity in news, or the lack thereof, political news is only one part of the solution. It is important to consider a broad range of topics. To investigate how social cohesion is affected, different genre need to be considered. The rising trend of both intentional and unintentional news avoidance is worrisome [37]. It can be motivated by content, the (technological) medium used, or the need for a "media detox" after an overload of information [3]. While political news is especially relevant for democratic discourse, all types of news need to be considered in diverse news recommendations.

**Recommendations.** Concretely, this results in a set of specific recommendations for scholars, practitioners, and regulators:

- The research community and regulators need to collaborate with industry to test the long-term impact of diverse news recommendations on relevant outcome variables that include not only democratic skills, like political knowledge. The wider societal impact needs to be investigated, such as fragmentation, radicalization, and polarization as well as the economic consequences for news providers and user experience and satisfaction. Research needs to move beyond investigating short-term effects, focusing on the long-term impact of the dynamic changes of how people interact with recommender system as well as their understanding thereof.
- Social scientists, computer scientists, and legal scholars need to work together and with journalists, regulators, and technology platforms to form a shared understanding of which types of diversity are important in the news domain, and why. This is crucial for making decisions on how to build such systems, and what should be optimized. An understanding of the goals that the algorithms are optimized for is also crucial in making decisions on the grounding of complex abstract concepts into technical definitions that are measurable and yield to optimization.
- All actors involved in designing, governing, and running news recommender systems should acknowledge that diversity is not the only absolute value, but needs to be balanced with other public values, such as privacy, autonomy, non-discrimination, and economic goals.
- System designers and social scientists should collaborate to better understand the way that users engage with and appreciate recommendations.
- Regulators and industry should collaborate to create an economic environment in which the formalization of diversity does not become a race to the bottom. Online news needs to be (self-)regulated in such a way that the societal responsibility ingrained in traditional journalistic organizations is not pressed out of the market.

---

[2] Operationalization is the process of making a concept like relevance of a news-item or diversity measurable. It is thus a precondition to any automatic decision making on the selection and order of digital content.

## 3 The user perspective on diversity: the tension between autonomy, diversity, and privacy

As the discussion so far highlights, diversity of perspectives and diversity of news in general is thought to be a cornerstone for the informed citizen in many democratic theories [22]. As such, the societal interest in informed citizens needs to be balanced against the individual interests of autonomous decision making. For the autonomous user, agency in a news recommender and control of its mechanics might result in an increased or reduced diversity of news. This may potentially create a tension between the editorial intention to diversify the news recommender and individual autonomy. Furthermore, a data-driven news recommender system, which might be used to cater to the autonomous user or/and to create a diverse news exposure, can be in conflict with privacy rights. Finally, there are economic goals to be protected, which might restrict the transparency and control that content providers, platforms, and news recommenders want to hand over to the user. The tension and balance between these values, and their relationship to each other, is under-researched. This leads to the following recommendations:
- Funding agencies should help establish frameworks for more open and transparent collaboration with platforms and the digital news industry to disentangle this complex network of conflicting interests and values.
- Regulators and research funders should ensure that more research can be done to better understand user behavior in *the broader information ecosystem* – both using traditional investigation approaches and novel ones such as active in-situ experimentation. It is important to not only consider *the wider media consumption landscape* (multi-platform, both editorial and social media content), but also the long-term effects on the relationship between news providers and the evolution of users' expectations and needs.

- The research community and industry should explore richer user models including both explicit and implicit elements. So far, research both in academia and in the industry primarily construes the user by the clicks they generate. The interest of the user as a citizen cannot be measured this way.

Following these recommendations would provide the foundation to address core user-oriented challenges and opportunities in the context of news recommender systems. In the following, we illustrate why and how research of the intricate network of relationships between these different values could inform policy decisions with respect to diversity, user autonomy, agency, transparency, and control.

**New modes of interaction between news generators and audience.** News recommender systems change the way users and the news media interact. On the one hand, news recommender systems are giving the media new tools to bring information to the attention of users and steering audience behavior. On the other hand, news recommendations can, at least in theory, be a means to bring journalists one step closer to the goal of being more responsive to the information needs (such as developing expertise in a topic, broadening horizons, or exploring the unexplored) and preferences of users, and truly engaging with the audience to build deeper, more fruitful relationships.

We therefore recommend that funding agencies initiate research programs which explore new ways of communication between media and audience, by, for example, improving transparency, user agency and control, and other feedback channels.

**User agency in a dynamic system.** A news recommender system is very often a dynamic system in which direct user choices as well as indirect traces derived from behavior feed into the algorithm and partially define the scope and diversity of future choice options, which in turn leads to new user choices. If a reader, for example, expresses a preference for news about crime and also consumes a lot of information on crime, this will affect the future choices made available by a news recommender. Assuming the news recommender only optimizes recommendation accuracy based on prior behavior, as is often the case today, then this likely leads to a de facto reduction in the diversity of news offered to this consumer. Improved interface design can both allow and encourage the user to provide explicit feedback on the recommendations offered and change preferences accordingly.

Proper interface design of news recommenders can thus support the user to become more aware of the system's inner workings and its influence on their media consumption. Interface design should make clear which defaults or presets the system has (e.g., initially applying a certain algorithm, even if users can change this in a control panel), so the user can decide whether or not recommendations are in line with their preferences. More engaged users can change the default settings, while default settings could be set towards nudging the user to be more diverse.

- We recommend that funding agencies put an emphasis on research into user agency and user-recommender interaction. Given the importance of interface designs and the adaptive nature of these systems, users need insights into how a setting will likely change the output, that is, a user-centric diversity requires transparency as well as control. However, the factors that influence the type of transparency and control users benefit from are not yet well understood.
- Furthermore, we recommend that more research is directed at the area of how recommendations are influenced by the user base as a whole. Recommendations are not only dependent on the individual's choices or the choices of the news publisher, but also on the choices of others interacting with a news recommender. Research on understanding the influence of and interactions between the user base as a whole and the news publishers with the individual user's preferences is lacking, but knowledge about the collective nature and complex behavior of these systems is important to inform a new generation of news recommender systems that are aligned with societal goals while preserving individuals' rights.

Transparency is another important point to consider when talking about user agency. News recommender systems tend to distribute news from a large set of sources, some of which are largely unknown to readers. This may make it difficult for readers to understand the perspective and context from which a piece of news is written. To make an informed decision on how to understand any piece of information, the recommendation and its basis (e.g., rationale and collected data) should be made transparent to the users. This includes, as a minimum basis, some reasoning around why users are presented with a particular recommendation. Informed news consumption by a news recommender system can be further improved by also giving users control over their experience. In fact, when users understand how social and personalized recommendations work, they are more likely to engage and think carefully about that information [8].

Existing work indicates that the combination of transparency and control can also increase trustworthiness (while the results for transparency alone are weaker). We thus relate user agency to both transparency and control. While a notable amount of work about transparency and control in recommender systems has been conducted to date, there has been limited work of this nature in the domain of news recommendation [26, 20].

- We recommend that research agencies fund more research on the interrelation between user agency by transparency as well as control and its relation to diversity.
- We also recommend the development of research to understand when, how, and for whom different transparency and control mechanisms are effective.

## 4 Data access, design, and measurement

To render the concept of diversity amenable to computational approaches, social and computer scientists need to work towards quantifiable, meaningful, and relevant definitions of diversity and its related concepts, knowing that there cannot be a single one. The way the different understandings of diversity impact users' news consumption requires oversight and academic research. The primary desideratum is therefore an increased awareness, recognition, and support of diversity as one of the explicit goals in news recommender systems. Research and implementations benefit not only from insights into potential negative implications, but also from an explicit appreciation of the positive role that diversity may play in the optimization of existing goals.

Optimization criteria for diversifying recommendations come with a set of normative assumptions that need to be carefully considered. This includes the impact on society and possible risks generated by the system in question. System designers should explicitly take into account the desired and projected effects on society as well as the causal relationship between the chosen metrics and the types of diversity. As system assumptions and optimization targets might change over time, monitoring needs to be a continued process.

A key prerequisite to the functional understanding of diversity is the deliberate interdisciplinary creation of operationalizations, that is, the definition of a quantifiable measure for a social concept. This requires a growing willingness of computer scientists to discuss social theories and an increased readiness of social scientists to learn about technical aspects, such as the functionalities of recommendation algorithms. Based on such collaborations, interdisciplinary teams are then recommended to create combined, larger and much more meaningful annotated data sets that serve as a basis for better news recommendation systems.

We also strongly recommend that a unified commitment to ethical standards in the form of a code of conduct between the aforementioned stakeholders is established, safeguarding reasonable and responsible use of data access (see Section 5). Such a code would serve to reduce legal uncertainty, safeguard the rights of citizens as well as organizations, help shape the terms of use, and provide a strong foundation for the future development of the field. Ensuring adequate handling of protected minorities or certain viewpoints in particular may necessitate sharing of person-related sensitive data, in which case a well-defined process should be in place to prevent potential misuse.

The following box gives an example of why data access, metrics, and analytic tools are necessary to give insight into the important societal process of news selection, prioritization, and distribution.

> **Example:** An interdisciplinary research project strives to (automatically) assess whether news services expose citizens to a diverse set of viewpoints, for example, through a diverse representation of political positions. For this purpose, news organizations need to provide:
> - A set of news articles.
> - Existing metadata insofar as they exist, such as the political viewpoints expressed therein, in conjunction with information on the annotation method.
> - Representative user data, following fully transparent sampling methods, describing the interaction of users with the news articles (e.g., accessing, clicking, and commenting).
> - Data describing the exact presentation and context of news to users by news recommender systems.
>
> The data have to cover a sufficient span of time to reflect a general pattern and avoid biases due to extraordinary events. Multiple news organizations need to provide data covering the same time frame to generate a more representative spectrum of the news ecosystem. Researchers can then approximate exposure to viewpoints across platforms and over time, thus getting a better understanding of the role of news recommender systems on the diversity of news exposure.

## 5 Governance

As we see in this manifesto, diversity and its future pathway is contingent on emerging technologies and future recommender systems, but it is also shaped by changing social norms, usage practices, regulations, and business models. As a consequence, democratic societies have, over time, developed institutions and regulations that aim to balance the public fostering of diversity on the one hand with the goal of securing independent media on the other [25]. A governance perspective takes into account this interplay between technological developments, social norms and values, regulatory interventions, and market transactions [9].

In broadcasting regulation, diversity is a main goal in many countries [34]. These regulations aim to guarantee that the diversity of existing opinions is presented in broadcasting as broadly and comprehensively as possible (BVerfG, Judgment of March 25, 2014 – 1 BvF [ZDF]). For example, public service broadcasters are required to promote diversity, and media regulators regulate the distribution of broadcasting programs to maximize diversity.

The turn to today's fragmented media environments and the increasing role of recommender systems, both at the level of media organizations and at the level of platforms and aggregators, challenge such regulatory approaches in three dimensions: (1) What is the very object of diversity? (Options, viewpoints, and values, etc.) (2) What is the scope of diversity regulation? (3) Can a system of representation ("existing opinions in society represented in a program/recommender system") still work?

Regulatory concepts reach their limits if diversity regulations are applied to media in an information ecosystem when media content is just one among many types of content that similarly fulfil the information and leisure needs of users [36]. Thus, meaningful regulation needs to go back to basic functions, such as ensuring free and open public and individual opinion formation and exchange (see Section 2).

We recommend that both regulators and legislators, as well as stakeholders in this sector, such as news media and platforms, take careful, evidence-based measures to guarantee diversity in news and opinion in future media and news environments.

**No direct transfer of broadcasting regulation.** The concept of diversity regulation in the domain of broadcasting should not be transferred to information ecosystems where media content is only one type of content that fulfills the information needs of users. If there is a need to regulate at all, the concepts should focus on specific obstacles to the openness of public communication (e.g., due to monopolistic structures giving actors undue power) or the options for the user to choose (e.g., between various recommender systems, see Section 3).

**Media and platforms to take on responsibility.** Media has always had the constitutive function to supply society with diverse and truthful information and opinions for democratic deliberation (see Section 1 and 2). Journalism has developed routines and institutions to guarantee this. Platforms are now partly taking over this role by mediating large shares of news and information consumption. Thus, both journalism and platforms now need to take on this shared responsibility. News organizations are in a position to leverage their century-long expertise in balancing commercial interests and democratic functions, but need to more systematically bring together this journalistic expertise with technological development. For platforms, this means taking responsibility for the impact they have on diversity, including opening up data and application programming interface for researchers to audit recommender systems and their impact (which requires overthinking policies on contracts and trade secrets). Furthermore, platforms need to allow users to exert agency (see Section 3) by exercising influence on news recommendation algorithms, and prioritize the media sources they would like to see.

**Code of conduct for data sharing.** Since academic research for the public good in this field depends to a large extent on data from the industry, we need to find a way to exchange data and keep the use of data conforming with legal regulations, especially GDPR. Article 40 of the GDPR allows for the drafting of codes of conduct. Such a code of conduct – after having gone through a specific procedure – becomes a binding governance instrument for the domain it covers. We recommend that the research community and platforms jointly draft a respective code that specifies legal modes of data sharing and establishes responsibilities for both platforms and researchers.

**Promoting R&D for legacy media.** To protect freedom of communication, self-regulation is the preferred choice in the field of public communication (which does not exclude legal regulation in certain cases). Legacy media and social media platform providers alike need to be aware of the consequences of their editorial and design choices in terms of diversity, and govern themselves responsibly. However, traditionally, media companies did not engage much in R&D and so funders should support them in this field. Given their experience in producing and distributing diverse content, public service media could be socio-technical innovators in the field of news recommender systems and should be given the mandate to develop, implement, and share new systems.

# 6 A new initiative for diversity in news recommendation

This manifesto addresses the fundamental developments in the production, dissemination, and usage of news as well as the role recommender systems have in these interrelated contexts. These changes challenge the traditional concept of diversity used in the media industry, research, and governance. In the context of the fragmented media ecosystem, we need to revisit the fundamental functions of diversity in society, and the impact of news recommender systems on society.

Specifically, we can summarize the considerations in the previous sections of this manifesto into the following high-level observations and put forward a list of recommendations to frame future initiatives.

**Cross-cutting observations.** The combination of insights from computer science, social science, and legal sciences in the Dagstuhl Perspectives Workshop 19482 have made the following clear:

- New measures and models of diversity are needed as current models of diversity, typically, do not capture the multidimensionality of diversity.
- The role of users and their ability to make choices about and have control over the news recommendation process is not sufficiently understood and attended to.
- The responsibility of news companies and platforms for supporting user agency should ensure transparency and user control.
- The data and models currently used in both computer and social science are often too restricted in terms of representativeness, duration, and depth to reflect the complexity of diversity as a societal concept.
- Governance in this space is opaque, but there are good reasons *not* to import the broadcasting paradigm of regulation and rather promote diversity through other means of regulation that ensure an ecosystem of diverse recommender systems.
- The joint investigation of the topics necessitates an interdisciplinary approach, where computer scientists are prepared to discuss social and legal theories, social scientists show an interest in technical and legal regulatory mechanisms, and legal scholars engage with technical and social mechanisms.

As a consequence, the workshop highlighted that in order to get valid answers to many of these pressing questions and challenges, *long-term, interdisciplinary research, including the fields of computer science, social science, economics, and the legal sciences, is needed.*

**Recommendations for future initiatives.** While our recommendations address several key issues in the application, development, and scientific study of news recommender systems, there are several questions that reach beyond the improvement of existing approaches and point to a need to consider news recommender systems in a wider context of socio-technical changes. Given the increasing blurring of boundaries of what constitutes news and how it is disseminated across platforms, several considerations discussed above may need to be extended beyond the realm of classic editorial content and distribution platforms. Societal

**A. Bernstein, C. de Vreese, N. Helberger, W. Schulz, K. Zweig, et al.**          **57**practices such as news avoidance raise new questions concerning the management of diversity reaching beyond those venues where users expect to encounter news. Investigating the dynamics of news recommendation, consumption, and opinion formation, as well as the positive and negative consequences of targeting and personalization, may call for entirely new research approaches.

To address all these challenges requires a *collaboration between disciplines* and *the development of new, interdisciplinary grounded measures of various diversity types* that are aligned with the societal goals and are concrete enough that they lend themselves as a computational implementation. More specifically, the following initiatives are needed:

**Conduct interdisciplinary research on news recommenders and diversity.** As most pressing societal and scholarly questions about news recommender systems and diversity cannot be answered meaningfully from a mono-disciplinary perspective, we call upon the (inter)national research community to organize and engage in truly interdisciplinary, continuously cooperating communities across computer, social, and legal sciences.

**Create a safe harbor for academic research with industry data.** Much research on public communication and recommender systems requires access to industry data to produce results that are meaningful for society. To enable this, data protection issues must be resolved. We recommend creating a code of conduct under Article 40 of the General Data Protection Regulation (GDPR) to give this kind of data sharing a solid legal basis.

**Strengthen the role of public values in news recommenders.** News recommenders can be powerful tools to help users find their way in the plethora of available news, shape public opinion, and serve as a foundation for public cohesion. They are extensions of the traditional editorial task. Hence, we recommend that they should not just maximize for clicks and short-term revenue, but, mindful of the democratic function of the media, also strengthen public values that align with the overall mission of a news outlet.

**Create a meaningful governance framework for news recommenders.** While we see no fruitful way of transferring existing regulations from broadcasting to news recommenders, we recommend that regulators and legislators support the research required to build diversity-aware recommender systems and actively foster an environment that allows for the co-existence of multiple recommender systems and their preconditions. Such initiatives should be evidence-based.

**We recommend founding a joint lab to spearhead the needed interdisciplinary research, boost practical innovation, develop reference solutions, and transfer insights into practice.** This initiative and its lab must combine the best (inter)national expertise from fields like computer science, social and behavioral sciences, political philosophy, and law, as well as industry and regulators to ensure diverse, transparent, explainable, and fair news recommendations.

These discussions should not be conducted in isolation, but with partners from the civil society, government and administration, industry, as well as science. The following list of actors should be taking actions to follow up on our recommendations: [3]

**Governmental organizations** such as the
- European Digital Media Observatory (`https://edmo.eu/`)
- European Commission Directorate-General Connect
- European Regulators Group for Audiovisual Media Services (ERGA)
- National regulators (via ERGA)
- Council of Europe – CDMSI Steering Committee on Media and Information Society

---

[3] Note that this list does not claim to be complete.

**19482**

**Funding agencies** such as the
- European Research Council
- National research councils and agencies

**Industry associations** such as
- European Broadcasting Union (EBU)
- Association of Commercial Television in Europe (ACT)
- EuroISPA

**Scientific actors** such as
- European Communication Research and Education Association (ECREA)
- International Communication Association (ICA)
- The Association for Computer Machinery (ACM)
- National Computing/Informatics Associations
- National Research Associations
- National Research Academies

**Civil society actors** such as
- AlgorithmWatch
- Consumer protection unions

## 7 Participants

- Christian Baden
  The Hebrew University of Jerusalem, IL
- Michael A. Beam
  Kent State University, US
- Abraham Bernstein
  Universität Zürich, CH
- Claes de Vreese
  University of Amsterdam, NL
- Marc P. Hauer
  Technische Universität Kaiserslautern, DE
- Lucien Heitz
  Universität Zürich, CH
- Natali Helberger
  University of Amsterdam, NL
- Pascal Jürgens
  Johannes Gutenberg-Universität Mainz, DE
- Christian Katzenbach
  Alexander von Humboldt Institute for Internet and Society, DE
- Benjamin Kille
  Technische Universität Berlin, DE
- Beate Klimkiewicz
  University Jagiellonski, PL
- Wiebke Loosen
  Leibniz Institute for Media Research - Hans-Bredow-Institut, DE
- Judith Moeller
  University of Amsterdam, NL
- Goran Radanovic
  Max Planck Institute for Software Systems, DE
- Wolfgang Schulz
  Universität Hamburg, DE
- Guy Shani
  Ben Gurion University, IL
- Nava Tintarev
  Technische Universiteit Delft, NL
- Suzanne Tolmeijer
  Universität Zürich, CH
- Wouter van Atteveldt
  Vrije Universiteit Amsterdam, NL
- Sanne Vrijenhoek
  University of Amsterdam, NL
- Theresa Zueger
  Alexander von Humboldt Institute for Internet and Society, DE
- Katharina Zweig
  Technische Universität Kaiserslautern, DE